\documentclass[iop,apjl,twocolumn]{emulateapj}

\usepackage{amsfonts}
\usepackage{amsmath}
\usepackage{amssymb}
\usepackage{bm}
\usepackage{dcolumn}
\usepackage{graphicx}
\usepackage{graphics}
\usepackage[latin1]{inputenc}
\usepackage{latexsym}
\usepackage{rotating}
\usepackage[colorlinks=false]{hyperref}
\usepackage[all]{hypcap} 
\usepackage{xspace} 
\usepackage[usenames]{color}
\usepackage{mathrsfs}

\usepackage{ulem}
\definecolor {darkgreen}{rgb}{0.2,0.7,0.2}

\newcommand\be{\begin{equation}}
\newcommand\ba{\begin{eqnarray}}
\newcommand\ee{\end{equation}}
\newcommand\ea{\end{eqnarray}}
\newcommand{\bes}{\begin{subequations}}
\newcommand{\ees}{\end{subequations}}
\newcommand{\beqn}{\begin{eqnarray*}}
\newcommand{\eeqn}{\end{eqnarray*}}
\newcommand\bw{\begin{widetext}}
\newcommand\ew{\end{widetext}}

\newcommand{\eff}{{\mbox{\tiny eff}}}

\begin{document}
\title{Spin-Precession: Breaking the Black Hole--Neutron Star Degeneracy}

\author{Katerina Chatziioannou}
\author{Neil Cornish}
\author{Antoine Klein}
\author{Nicol\'as Yunes}
\affiliation{Department of Physics, Montana State University, Bozeman, MT 59717, USA.}

\begin{abstract} 

Mergers of compact stellar remnants are prime targets for the LIGO/Virgo gravitational wave detectors. 
The gravitational wave signals from these merger events can be used to study the mass and spin distribution of stellar remnants,
and provide information about black hole horizons and the material properties of neutron stars. However, it has been suggested that
degeneracies in the way that the star's mass and spin are imprinted in the waveforms
may make it impossible to distinguish between black holes and neutron stars. Here we show that the precession of the
orbital plane due to spin-orbit coupling breaks the mass-spin degeneracy, and allows us to distinguish between
standard neutron stars and alternative possibilities, such as black holes or exotic neutron stars with large masses and spins.
\end{abstract}

\maketitle

\section{Introduction}

 Compact stellar remnant mergers are the main targets of gravitational wave (GW) detectors such as advanced LIGO (aLIGO)~\citep{Harry:2010zz} and advanced Virgo (Adv)~\citep{Acernese:2007zze}, with predicted rates between a few and a few hundred per year at full design sensitivity~\citep{Abadie:2010cf}. These systems 
take tens of minutes to sweep through the sensitive band of the detectors, entering the band at $\sim10$Hz, and terminating in the kHz range with a violent merger lasting just a few milliseconds.

The final stages of the inspiral and merger proceed differently for black holes (BHs) and neutron stars (NSs), and in principle, this should allow us to identify the
make-up of the system from the GW signal alone. However, the number of GW cycles in the signal and the aLIGO/AdV sensitivity fall off rapidly with increasing frequency, 
meaning that there is very little information past $\sim 500$ Hz (less than 2\% of the SNR).  Probes of BH physics and the equation of state of NSs will likely require multiple detections~\citep{DelPozzo:2011pg, DelPozzo:2013ala}. An electromagnetic counterpart to the GW signal, such as a short-hard gamma-ray burst or an associated kilonova/macronova emission~\citep{Metzger:2011bv},
would indicate that at least one of the bodies was a NS, but beaming effects or the luminosity of the signal may make detecting a counterpart difficult for the majority of mergers~\citep{Abadie:2010cf,Aasi:2013wya}.
Absent a counterpart, we must rely on the early inspiral to extract information about the make-up of the binary, which poses a challenge since finite size effects are completely negligible during inspiral~\citep{Read:2009yp}. All we have to go on to decide the composition of the binary are the values of the masses and spins inferred from the inspiral signal.

General arguments based on stability and causality limit the mass and spin of NSs to the range $M\in[0.1, 3.2 ]M_\odot$ for the mass and $\chi \in[0,0.7]$ for
the dimensionless spin magnitude,~$\chi \equiv |\vec{S}|/M^{2}$, where $\vec{S}$ is the spin angular momentum~\citep{Rhoades:1974fn,Lattimer:2006xb,Yagi:2014bxa}. Realistic equations of state yield a tighter mass range $M\in[1.0, 2.5] M_\odot$. The observed range of masses and spins is
somewhat tighter~\citep{Lattimer:2006xb,Ozel:2012ax}: $M\in[1.0, 2.0]M_\odot$, $\chi \in[0,0.3]$. The old NSs that merge are expected to have spun down by magnetic breaking to the point where
the maximum spin is much lower, $\chi \lesssim 0.05$, than in the general NS population~\citep{Mandel:2009nx}. Furthermore, the standard isolated NSNS binary formation scenario ensures that after every common envelope phase (that tends to align the spins) follows a supernovae kick that misaligns the spins (unless the kick is in the orbital plane, though there is evidence that this is not the case~\citep{Kaplan:2008qm}). Thus, we adopt the definition that {\em normal} NSs seen
by aLIGO/AdV have $M\in[1, 2.5]M_\odot$ and $\chi \leq  0.05$, and term NSs with larger masses or spins {\em exotic}. Einstein's theory of gravity allows BHs to have spin in
the range $\chi \in[0,1]$ with any mass. X-ray observations have identified stellar remnant BHs with $M\in [3.6, 36]M_\odot$ and $\chi \in[0,1]$. There is currently some debate as to the existence of a mass gap between NSs and BHs~\citep{Ozel:2010su,Farr:2010tu,Belczynski:2011bn}, but for the purpose of determining whether a normal NS could be misidentified as a BH or an exotic NS, the existence of a gap is moot.

The early inspiral phase of a compact binary merger can be modeled analytically by expanding Einstein's equations in powers of the ratio of the orbital velocity to the speed of light, the so-called \emph{post-Newtonian} (PN) approximation~\citep{Blanchet:2013haa}. This ratio is small during the inspiral, with $v/c$ of $7\%$ when the system enters the detector sensitivity band, reaching roughly $40\%-60\%$ by contact~\citep{Bernuzzi:2014kca}. The PN approximation becomes less accurate as the system evolves through the band, eventually breaking down at the end of the inspiral phase. As all forms of energy couple to gravity, both the masses and spins leave an imprint on the binary orbit and the GWs emitted. The coupling between spin and orbital angular momentum can strongly affect the orbital trajectory and the GWs emitted in the inspiral phase. 

The PN approximation can be used to construct a model of the GWs emitted during inspiral. The combination of such a GW model with a model for the instrument response yields templates for the signals as seen by the detector. Subtracting the model from the data produces a residual, and demanding that the residual is consistent with a model for the instrument noise defines a likelihood function. From this function and our prior knowledge we can derive a posterior distribution for the model parameters that are consistent with the observed data. It often happens that there are strong correlations between these parameters, limiting our ability to measure each parameter individually.

Recent work~\citep{Hannam:2013uu} has suggested that the correlation between mass and spin~\citep{Cutler:1992tc,Cutler:1994ys} may make it impossible to distinguish between a NSNS binary and a NSBH or a BHBH binary. This result hinges on a simplified waveform model that assumes that the spin and orbital angular momenta are perfectly aligned, and thus, spin-orbit induced precession~\citep{springerlink:10.1007/BF00756587,Bohe:2012mr} is absent. However, we have no reason to expect the spin and orbital angular momenta to be aligned in stellar remnant binaries. Indeed, the NS binaries observed at much longer orbital periods are far from aligned and are precessing~\citep{2002ApJ...576..942W,2005ApJ...624..906H,2008Sci...321..104B}. It has been hypothesized~\citep{Hannam:2013uu,Baird:2012cu} that spin precession would not significantly alter the conclusions drawn using spin-aligned waveforms. We have tested this hypothesis and found, as first suggested by~\citet{Cutler:1992tc}, that spin precession adds additional richness to the signals that almost completely breaks the mass-spin degeneracy,  producing an order-of-magnitude improvement in the extraction of the individual masses and spins, which allows us to distinguish between NSs and BHs.
We show that normal NS binaries will not be mistaken for BHs or exotic NSs, but we cannot rule out the possibility that some exotic NSs or low mass/low spin BHs may
be misidentified as normal NSs.

\section{Methodology}

We employ Bayesian inference~\citep{Trotta:2005ar,Cornish:2007ifz,Littenberg:2009bm,Aasi:2013jjl} to quantify the astrophysical information can be extracted from a GW detection. In particular, when comparing models, we compute the so-called \emph{Bayes Factor} (BF), which is the ratio of the evidence for one model to that for another. We compute BFs through Markov-Chain Monte-Carlo (MCMC) techniques, as described by~\citet{Cornish:2007ifz,Littenberg:2009bm,Aasi:2013jjl}, with the high-power, zero-detuned noise spectral density of aLIGO and AdV~\citep{AdvLIGO-noise}. We consider only the inspiral phase, from $10$ Hz up to $400$ Hz, at which point NS tidal deformations can no longer be neglected; extending the analysis beyond $400$Hz would only strengthen the results obtained here. With these tools, and assuming a GW detection, we address the following questions:
\begin{enumerate}
\item Can we distinguish between NSNS binaries and low-mass, small-spin NSBH binaries only by the inspiral portion of the waveform?
\item Can we distinguish between non-spinning and spinning binaries with the data only?
\item Is the mass uncertainty large enough to lead to a false detection of astrophysically ``exotic" NSs?
\end{enumerate}

To answer these we need a waveform template that accurately models the GWs emitted during the quasicircular inspiral of spin-precessing, compact binaries. Previous studies were limited to spin-aligned or antialigned systems~\citep{Hannam:2013uu,Baird:2012cu}, as until recently, these were the only systems for which fast, closed-form frequency domain waveforms were available (numerical time-domain templates are available, but their high computational cost limits their utility, see the discussion in~\citet{Chatziioannou:2014bma}). Recently, analytical models for precessing systems were derived by noting that in the inspiral phase, three intrinsic scales separate: the orbital timescale is shorter than the precession timescale, which is shorter than the radiation-reaction timescale~\citep{Hinderer:2008dm,Klein:2013qda,Chatziioannou:2013dza} (see also~\citet{Lundgren:2013jla}). This separation allows us to solve the PN precession equations analytically through a perturbative expansion about small spins and multiple-scale analysis. Once the orbital motion has been computed, the Fourier transform of the waveform can be constructed through the stationary-phase approximation~\citep{Droz:1999qx,Yunes:2009yz}, leading to small-spin~\emph{double-precessing} templates. 

The usefulness of any waveform template hinges on its accuracy relative to the true signal. In~\citet{Chatziioannou:2013dza}, we compared these double-precessing templates to numerical evolved PN waveforms. We found that the double-precessing model is highly accurate for all plausible NS spin magnitudes, however it fails for systems with BHs that possess large spins and precess significantly~\citep{Chatziioannou:2014bma}. We found that the integrated cross-correlation (the \emph{match}, sometimes called the \emph{faithfulness}~\citep{Damour:1997ub}) is above the $98\%$ [Fig.~(1) of~\citet{Chatziioannou:2013dza}]. For systems with SNR$ \lesssim 20$, this implies that the statistical error dominates over the systematic error (see Appendix 1 of~\citet{Chatziioannou:2014bma} for a proof of the independence of statistical and systematic errors). 

Each model incorporates spin effects in a different way and has, thus, a different spin prior. For the double-precessing model, we use uniform priors on the spin magnitudes and uniform priors on the unit sphere for the spin angles. For the spin-aligned model we again use uniform priors on the spin magnitudes, but delta functions about (anti)alignment with the orbital angular momentum for the spin angles. Clearly, the prior used in the double-precessing case is the most generic one since it assumes the least amount of prior information about the signal. Furthermore, a prior favoring spin alignment is not supported by astrophysical data. All the models use uniform priors on the masses.

\section{Distinguishing between NSs and BHs}

\begin{figure}[h]
\begin{center}
\includegraphics[width=\columnwidth,clip=true]{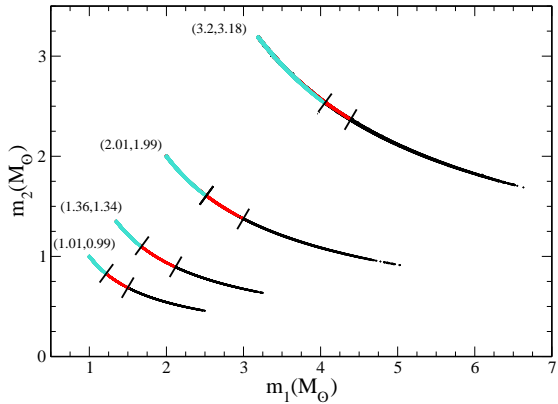}
\caption{\label{fig:m1m2_spin0} (Color Online)  Scatter plot showing points from the $90\%$ probability quantile in $(m_1,m_2)$ for non-spinning signals with different masses of SNR 10 extracted with non-spinning (turquoise), spin-aligned (black), and double-precessing (red) templates. The posteriors overlap from the equal mass boundary to the short lines that cut across the scatter plots indicating the separation between the different posteriors in the direction orthogonal to the chirp mass. The use of double-precessing templates leads to more accurate mass extraction.}
\end{center}
\end{figure}

We simulated four non-spinning systems with different masses and recovered them with non-spinning, spin-aligned~\citep{Poisson:1995ef,Arun:2008kb,Lang:2011je,Ajith:2011ec}, and the double-precessing models~\citep{Chatziioannou:2013dza}. All signals have a declination $\cos{\theta_N}=-0.11$, right ascension $\phi_N=3.71$, and inclination angle $\iota=63^{\circ}$ (all chosen randomly, requiring that they do not correspond to any special configuration, like optimal orientation or spin alignment). Figure~\ref{fig:m1m2_spin0} shows a 2D scatter plot of points in the $(m_1,m_2)$ plane (with $m_1\geq m_2$) that belong in the $90\%$ probability quantile of the posterior distributions. The points are clustered along lines of constant chirp mass, ${\cal M}=(m_1 m_2)^{3/5}/(m_1+m_2)^{1/5}$, where $m_{1,2}$ are the binary's component masses, with the extent of the lines determined by how well the dimensionless, symmetric mass ratio $\eta=m_1 m_2/(m_1+m_2)^2$ is determined. For waveforms with spin, the degeneracy between spin and mass ratio enlarges the $90\%$ confidence region.
\vspace*{-0.1in}

How well the mass ratio can be measured depends on the particular model used. Non-spinning templates lead to the smallest spread in the recovered masses, but at the cost of large systematic biases when one considers astrophysical realistic spin-pressing signals. Spin-aligned templates measure the mass ratio with a larger spread, due to degeneracies between masses and the spins. The inclusion of spin-precession partially breaks this degeneracy, translating into an improvement in the accuracy of the mass extraction that resembles what one would obtain with non-spinning templates. Similar results are shown in~\citet{Chatziioannou:2014bma} for signals with $\chi_{1,2} = 0.04$. 

\begin{figure}
\begin{center}
\includegraphics[width=\columnwidth,clip=true]{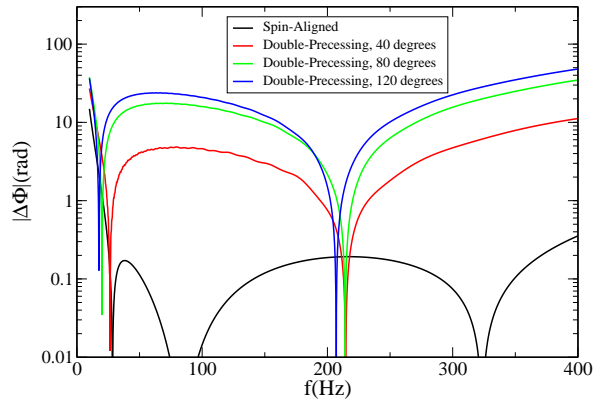}
\caption{\label{fig:dephasing} (Color Online) Phase difference between the nonspinning system $(m_1,m_2)=(1.36,1.34)M_{\odot}$ of Fig.~\ref{fig:m1m2_spin0} and a system that belongs in the $90\%$ probability quantile of the spin-aligned model with $(m_1,m_2)=(2.51,0.79)M_{\odot}$ and $(\chi_1,\chi_2)=(0.04,0.82)$(black line). Keeping the masses and the spin magnitudes of the second system fixed, we misalign the spins and plot the phase difference between the initial nonspinning system and the new precessing system for $40^{\circ}$ (red line), $80^{\circ}$ (green line), and $120^{\circ}$ (blue line) between the total spin and the orbital angular momentum at 10Hz. The dephasing induced by the spin-aligned model is below 1 radian for a wide range of frequencies $[20,400]$Hz, a manifestation of the mass-spin degeneracy. On the other hand, the double-precessing model results in a large dephasing, and hence a bad fit, which breaks the mass-spin degeneracy.}
\end{center}
\end{figure}

The leading order spin effects in the waveform enter through the effective spin parameter $\chi_{\eff} \equiv (\vec{\chi}_1 \cdot \hat {L} + \vec{\chi}_2 \cdot \hat {L})/2$, where $\vec{\chi}_{1,2} \equiv \vec{S}_{1,2}/m_{1,2}^{2}$, $\vec{S}_{1,2}$ is the spin angular momentum of the binary components and $\hat{L}$ is the unit orbital angular momentum. To check if the improved parameter estimation was due to the prior on
$\chi_{\eff}$ we performed an analysis with spin-aligned templates using the same prior on $\chi_{\eff}$ that was used for the spin-precessing model and found that the results are not altered~\citep{Chatziioannou:2014bma}. The explanation lies in the likelihood, not the prior: the extra freedom in the spin orientation of the precessing model makes it less likely for systems with large masses or spins to match the signal. 
Figure~\ref{fig:dephasing} illustrates this through the dephasing between one of the systems of Fig.~\ref{fig:m1m2_spin0} and a system whose mass and spin magnitude are in the $90\%$ probability quantile of the spin-aligned model (but not in that of the precessing one), for different angles between the total spin and the orbital angular momentum. The spin-aligned system induces a very small dephasing despite the high value of $\chi_2$, indicating the presence of a mass-spin degeneracy. On the other hand, the double-precessing systems results in large dephasings, leading to a low likelihood.

The fact that even a very small transverse spin can have such a big effect on data analysis can be understood as follows. Spin-alignment introduces a very strong correlation between the masses and the spins. As a result, the parameter covariance matrix is near singular. The near singularity of the covariance matrix means that very small changes in the waveforms can have significant effects on parameter estimation. Even the small amount of precession expected for NS binaries is sufficient to alter the mass-spin correlation and lead to very different parameter estimation results. 

By breaking the degeneracy between masses and spin magnitudes one obtains higher accuracy in the extracted masses, which in turn implies one would be able to distinguish between NSs and low-mass, small-spin BHs. This is not the first time that the inclusion of spin-precession in the templates has been shown to improve parameter extraction dramatically~\citep{Vecchio:2003tn,Lang:1900bz,PhysRevD.80.064027}, relative to spin-aligned templates~\citep{Lang:2011je}. For example, projections for the bounds on the mass of the graviton and the Brans-Dicke parameter using spin-aligned templates~\citep{Berti:2004bd} were up to an order of magnitude larger
than those found for non-spinning systems~\citep{Will:1994fb,Scharre:2001hn,Will:1997bb,Will:2004xi}. Including spin-precession effects~\citep{Stavridis:2009mb,Yagi:2009zm} broke parameter degeneracies
and gave projected bounds similar to those for non-spinning systems.

\section{Distinguishing between non-spinning and spinning binaries}

Before we can discuss distinguishability between spinning and non-spinning systems, we must understand how spin enters the waveform templates. For systems with similar component masses, spin first enters through the effective spin parameter  $\chi_{\eff}$. Not surprisingly, this is the parameter that can be extracted most accurately, just like the chirp mass is measured more accurately than the symmetric mass ratio. In this case, however, a measurement of $\chi_{\eff}$ only provides information about the component of the spin angular momentum along the orbital one. Measuring the perpendicular components of the spin angular momentum would require measuring the cone of precession, which is difficult with the SNRs expected with aLIGO.

\begin{figure}[t]
\begin{center}
\includegraphics[width=\columnwidth,clip=true]{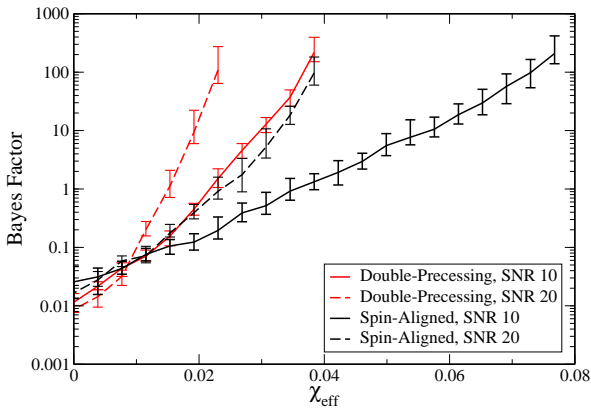}
\caption{\label{BF_spin} (Color Online) BF as a function of $\chi_{\eff}$ between non-spinning and spinning models for spin-aligned (black) and double-precessing (red) templates, assuming a precessing simulated signal with SNR $10$ (solid) and $20$ (dotted) and $(m_1,m_2)=(1.43,1.23)M_{\odot}$ in an aLIGO-AdV network.}
\end{center}
\end{figure}

We tackle the distinguishability of spinning and non-spinning systems as a model selection problem~\citep{Trotta:2005ar,Cornish:2011ys,Gossan:2011ha,DelPozzo:2011pg,Sampson:2013lpa}. We use a precessing system, with the total spin angular momentum vector at $30^\circ$ from the orbital angular momentum, and the same sky location used in Fig.~\ref{fig:m1m2_spin0}. We recover this signal with either the spin-aligned or the double-precessing model. Figure~\ref{BF_spin} shows the BF between non-spinning and spinning models as a function of the $\chi_{\eff}$ for signals with SNRs of $10$ and $20$. ${\rm{BF}}>1$ indicates that the data favors the spinning model. For the same SNR the double-precessing template correctly identifies the signal as produced by a spinning source at a lower value of $\chi_{\eff}$ than the spin-aligned model, while both models correctly identify a non-spinning signal ($\chi_{\eff}=0$ case). The details of the calculation of the BF are presented in~\citet{Chatziioannou:2014bma}. 

\section{Distinguishing between normal and exotic NS binaries}

Imagine we have detected a GW produced by a NSNS binary. The double-precessing model has enabled us to correctly identify it as consisting of NSs. But, are the remaining parameter uncertainties enough to lead to an erroneous inference that we have detected a NS with parameters outside those expected from astrophysical models? We define a \emph{normal} NS binary as one with $m_{1,2} \in [1,2.5]M_{\odot}$ and $\chi_{\eff} \in [-0.05,0.05]$ and an \emph{exotic} NS as one that is not normal. We could have chosen different values for the boundaries in $m_{1,2}$ and $\chi_{\eff}$, but these are consistent with current astrophysical considerations, and the results would not qualitatively change if we picked other values. Notice that a $\chi_{\eff}$ in that range does not guarantee $\chi_{1,2} \leq 0.05$, due to the effect of the projection along the orbital angular momentum. Nonetheless, a detection of a system with $\chi_{\eff} \geq 0.05$ would unambiguously imply that the system possesses at least one $\chi \geq 0.05$.
\begin{figure}[htb]
\begin{center}
\includegraphics[width=\columnwidth,clip=true]{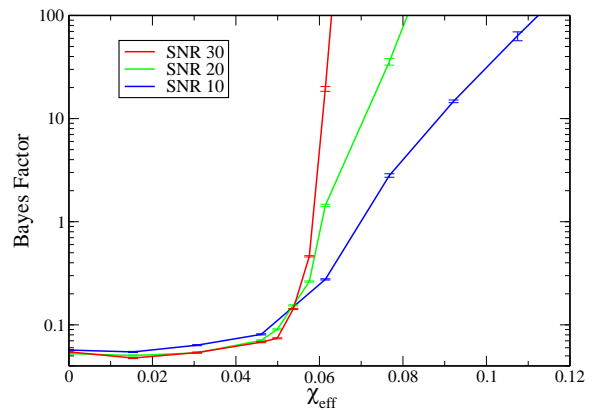}
\caption{\label{BF_exotic} (color online) BF in favor of the exotic NS model as a function of $\chi_{\eff}$ for different SNR values. The simulated signal is precessing and $m_1=1.43 M_{\odot}$ and $m_2=1.23 M_{\odot}$.}
\end{center}
\end{figure}

Figure~\ref{BF_exotic} shows the BF in favor of the exotic NS model for a precessing signal as a function of $\chi_{\eff}$ for different SNRs using the double-precessing model. Regardless of the SNR of the signal, a normal NS is always recovered as such. There exists, however, a window in parameter space (signals with SNRs of $10$ and $\chi_{\eff} \in [0.05,0.07]$) that could lead to the characterization of the system as normal, when in reality it was exotic. 

\section{Conclusions}

We showed that the inclusion of spin-precession in waveform templates breaks the degeneracy between the system's individual masses and spins, and allows us to distinguish between NSNS binaries and low-mass, small-spin NSBH or BHBH binaries. Moreover, even for signals with modest SNR, we can distinguish between ``normal'' and ``exotic'' NSs. These results open the door to population studies with the first GW detections, as well as coincident studies between the electromagnetic detection of short gamma-ray bursts and GWs. Indeed, if such a coincident observation is made, being able to identify the source from purely GW observations as a NS binary, a mixed binary or a BH binary would prove invaluable. 

The results presented here are subject to several assumptions. First, the noise is assumed to be stationary and
Gaussian, while in reality this may not be the case. Proper noise modeling along the lines described
in~\citet{Littenberg:2014oda,Cornish:2014kda} will help to restore performance to levels close to the ideal case.
Second, calibration errors and non-stationary drifts in the noise spectrum should be marginalized over in a full analysis, but these mostly impact the amplitude parameters, and only have a small impact on the spin measurement. Third, the waveform model inaccuracies do not affect our estimates of the statistical errors at leading order~\citep{Chatziioannou:2014bma}, so our conclusions will apply to more accurate waveform models.

\newpage 

\emph{Acknowledgments}.~We would like to thank Laura Sampson for many helpful discussions and Cole Miller for providing us with information about astrophysical expectations of NS spin orientations. We thank Thomas Dent, Mark Hannam, Richard O'Shaughnessy, and Evan Ochsner for comments and suggestions. K.~C.~acknowledges support from the Onassis Foundation. N.~Y.~acknowledges support from NSF grant PHY-1114374, NSF CAREER Grant PHY-1250636 and NASA grant NNX11AI49G. A.~K.~and N.~C.~acknowledge support from the NSF Award PHY-1205993 and NASA grant NNX10AH15G.  

\bibliographystyle{apj}


\end{document}